\documentclass[%
 reprint,
 amsmath,amssymb,
 aps,
]{revtex4-2}

\usepackage{graphicx}% Include figure files
\usepackage{dcolumn}% Align table columns on decimal point
\usepackage{bm}% bold math
\usepackage{algorithm}
\usepackage{algpseudocode}
\usepackage{booktabs}
\usepackage{braket}

\begin{document}

\preprint{APS/123-QED}

\title{QPP-RNG: A Conceptual Quantum System for True Randomness}
% Force line breaks with \\
\thanks{A footnote to the article title}%

\author{Randy Kuang}
  \email{randy.kuang@quantropi.com}
\affiliation{%
Quantropi (Canada)\\
 1545 Carling Av., Suite 620, Ottawa,ON K1Z 8P9, Canada. 
}%

\date{\today}

\begin{abstract}
We propose and experimentally demonstrate the \emph{Quasi-Superposition Quantum-inspired System (QSQS)} --- a conceptual quantum system for randomness generation built on measuring two conjugate observables of a permutation sorting process: the deterministic permutation count \(n_p\) and the fundamentally non-deterministic sorting time \(t\). By analogy with quantum systems, these observables are linked by an uncertainty-like constraint: algorithmic determinism ensures structural uniformity, while system-level fluctuations introduce irreducible unpredictability. We realize this framework concretely as \emph{QPP-RNG}, a system-embedded, software-based true random number generator (TRNG). In QPP-RNG, real-time measurements of sorting time \(t\) --- shaped by CPU pipeline jitter, cache latency, and OS scheduling --- dynamically reseed the PRNG driving the permutation sequence. This design fuses deterministic and non-deterministic components, so that entropy emerges organically from the quasi-superposition structure of the system. Crucially, QSQS transforms initially right-skewed raw distributions of \(n_p\) and \(t\) into nearly uniform outputs after modulo reduction. This effect arises from the system's internal degeneracies: many distinct internal states collapse into the same output symbol, effectively flattening biases and filling out the output space. This transformation from biased measurements to uniform randomness is the core principle of QSQS. Empirical results show that as the repetition factor \(m\) increases, output entropy converges toward theoretical maxima: Shannon and NIST SP 800-90B min-entropy values approach 8 bits, chi-squared statistics stabilize near ideal uniformity, and bell curve plots visually confirm the flattening from skewed to uniform distributions. Beyond practical implications, our findings illustrate how QSQS unifies deterministic algorithmic processes with non-deterministic physical fluctuations in a single framework, offering a physics-based perspective for engineering randomness. In the quantum-safe era, QPP-RNG can close the entropy gap by embedding true randomness directly into cryptographic modules, reducing reliance on external entropy sources and enabling entropy-rich, self-contained post-quantum cryptographic eco-systems.
\end{abstract}


\maketitle

%\tableofcontents
\section{Introduction}

Randomness underpins a wide range of scientific and technological domains, from statistical physics to modern cryptography and quantum information science~\cite{herrero2017physical, menezes1996handbook}. In cryptography, unpredictability is paramount: secret keys, nonces, and ephemeral randomness must exhibit both statistical uniformity and resistance to prediction~\cite{nist80090a}.

Random number generators (RNGs) are typically classified by their entropy sources: Pseudorandom Number Generators (PRNGs) deterministically expand a small initial seed~\cite{lehmer1949,xorshift,MacLaren1970,marsaglia2003xorshift}; True Random Number Generators (TRNGs) extract entropy from unpredictable classical physical processes~\cite{trng-ji-2020, trng-yang-2018, trng-ansari-2022, trng-tehranipoor-2023}; and Quantum Random Number Generators (QRNGs) exploit the intrinsic indeterminacy of quantum measurements~\cite{zhang2023qrng, qrng-Gabriel2010, qrng-Ma2016}.

This work originates from a generalization of the classical \emph{one-time pad} (OTP), whose perfect secrecy was established by Shannon in 1949~\cite{shannon1949communication}. Extending beyond Boolean algebra, we proposed the \emph{Quantum Permutation Pad (QPP)}~\cite{kuang2020qpp, qpp-springer-kuang-2022}, a lightweight yet powerful primitive formulated within a linear algebraic framework for quantum computing. QPP demonstrates striking versatility: it can be implemented on classical processors (via permutation matrices) or on quantum platforms (via permutation gates) defined over discrete Hilbert spaces~\cite{kuang2022epjqt, perepechaenko2023epjqt, Alain-qpp-runtime-2024, qpp-circuit-michel-2024}. Within this framework, the classical OTP appears as a degenerate special case, limited to just two permutations: identity and bitwise XOR.

Cryptographically, QPP dramatically enlarges the key space for \(n\)-bit data from \(2^n\) (Boolean algebra) to \((2^n)!\) (the full permutation group), while leveraging the generalized uncertainty of permutation operators to enable controlled key reusability~\cite{kuang2020qpp, qpp-springer-kuang-2022, kuang2025qpp}. Building on QPP, we developed the \emph{Homomorphic Polynomial Public Key} (HPPK) cryptosystem, which constructs key encapsulation mechanisms (KEMs)~\cite{kuang2023-HPPK-KEM} and digital signature schemes~\cite{kuang2024-hppk-ds-academia} using QPP arithmetic expressions. Collectively, these works position QPP as a unifying \emph{cryptographic DNA} spanning lightweight symmetric primitives, quantum encryption, and post-quantum asymmetric schemes~\cite{kuang2025qpp}, all framed within the broader theory of quantum cryptographic dynamics~\cite{kuang2025qcd}.

QPP further inspired \emph{pQRNG}~\cite{kuang2021qce}, a pseudorandom number generator designed to emulate cryptographically secure RNGs by leveraging the immense internal state space \((2^n!)^m\) formed by \(m\) independent permutations of an \(n\)-bit pad. While highly complex, pQRNG requires seeds of size \(m \times \log_2((2^n)!)\), beyond the practical scope of conventional PRNGs.

From this foundation emerged \emph{QPP-RNG}: a system-embedded, software-based True Random Number Generator that extracts \emph{physical entropy} from unpredictable microarchitectural fluctuations in commodity hardware—such as CPU pipeline timing variations, cache misses, memory access jitter, and OS scheduling effects. This raw entropy is then amplified through iterative permutations in QPP sorting cycles, producing high-entropy, statistically uniform random outputs entirely in software—without dedicated TRNG or QRNG hardware.

Empirical evaluation shows that QPP-RNG consistently passes rigorous statistical tests—including the NIST SP 800-90B Independent and Identically Distributed (IID) tests, the NIST SP 800-22 suite, and ENT tests—across diverse platforms ranging from general-purpose computing systems~\cite{qpp-rng-sci-kuang-2025} to mobile devices~\cite{kuang2025-qpp-rng-qinp}. Yet, despite this empirical robustness—and despite operating entirely on deterministic processors—a central scientific question remains: \emph{What physical mechanism explains QPP-RNG's ability to produce robust, IID true randomness?}

In this work, we propose a novel physical model: the \emph{Quasi-Superposition Quantum-inspired System (QSQS)}. We model the permutation sorting process as a deterministic evolution over a discrete basis $\{|0\rangle, |1\rangle, \ldots, |2^n-1\rangle\}$, while real-time system jitters, quantified by the sorting time \(t\), act as unpredictable measurements inducing state collapses. Within this hybrid system, the (modular) permutation count \(n_p\) and sorting time \(t\) emerge as \emph{conjugate observables}, mirroring the uncertainty principle in quantum mechanics. This perspective reveals how true entropy can arise from their interplay—even within classical computation.

By integrating entropy harvesting directly into the cryptographic primitive itself, QPP-RNG bridges both the \emph{logical entropy gap} (deterministic algorithms alone cannot create fresh entropy) and the \emph{physical entropy gap} (dependence on external TRNG or QRNG hardware). This integration strengthens security, reduces external attack surfaces, and enables practical, system-embedded TRNGs deployable on commodity devices.

The remainder of this paper is organized as follows: Section~\ref{sec:related} reviews related work on TRNGs, QRNGs, and OS-level entropy sources; Section~\ref{sec:model} formalizes the QSQS framework; Section~\ref{sec:experiments} presents cross-platform experiments and entropy convergence results; and Section~\ref{sec:conclusion} concludes.

\section{Related Work}\label{sec:related}

Random number generation spans classical and quantum domains, each grounded in distinct physical principles and operational methodologies.

\paragraph*{Pseudorandom Number Generators (PRNGs).}
PRNGs deterministically expand a short, high-entropy seed into long bitstreams~\cite{nist80090a, lehmer1949, xorshift, MacLaren1970, marsaglia2003xorshift}. While efficient, they cannot increase entropy beyond the seed, making unpredictability entirely reliant on seed secrecy and algorithm robustness.

\paragraph*{Classical TRNGs.}
True random number generators (TRNGs) extract entropy from practically unpredictable physical processes, including thermal noise~\cite{holman1997integrated}, avalanche noise~\cite{sunar2007true}, or metastability. A notable subclass exploits \emph{CPU jitter}—fine-grained variations from thermal, voltage, and microarchitectural effects. For instance, the JitterEntropy RNG~\cite{maurer2019jitterentropy}, integrated into the Linux kernel, measures these fluctuations via cycle counters.

\paragraph*{Quantum Random Number Generators (QRNGs).}
QRNGs derive entropy from fundamentally indeterministic quantum measurements, where superposed states collapse probabilistically according to the Born rule~\cite{herrero2017physical}. Implementations include single-photon detection~\cite{jennewein2000fast}, vacuum fluctuation sampling~\cite{gabriel2010generator}, and beam-splitter-based schemes~\cite{zhang2023qrng}. QRNGs offer strong theoretical guarantees, but typically require dedicated quantum hardware.

\paragraph*{Operating System Entropy Pools.}
Operating systems aggregate multiple entropy sources into deterministic random bit generators (DRBGs)~\cite{nist80090a}. Examples include Linux \texttt{/dev/random} and \texttt{/dev/urandom}~\cite{kern2014urandom}, Apple's \texttt{SecRandomCopyBytes}~\cite{apple2019security}, and Windows CNG~\cite{microsoftCNG}. Though practical, these systems remain fundamentally deterministic and depend on periodic reseeding.

\paragraph*{Prior Work on QPP and QPP-RNG.}
Earlier work introduced the \emph{Quantum Permutation Pad (QPP)}~\cite{kuang2020qpp, qpp-springer-kuang-2022}, which dramatically expands the key space by applying permutation operators instead of bitwise XOR. Building on this, \emph{QPP-RNG}~\cite{kuang2025axiv, qpp-rng-sci-kuang-2025} demonstrated that system-level jitter, when amplified through permutation sorting, can generate high-entropy randomness entirely in software.

\paragraph*{Current Contribution.}
This paper advances prior work by introducing the \emph{Quasi-Superposition Quantum-inspired System (QSQS)}, a physical model interpreting permutation count and sorting time as conjugate observables—mirroring quantum uncertainty—to explain QPP-RNG’s observed unpredictability and IID randomness.

\section{Theory and Model}
\label{sec:model}

In this section, we formally define the \emph{Quasi-Superposition Quantum-inspired System} (QSQS) and elucidate its role in the architecture of QPP-RNG. We then distinguish between two complementary operational modes: a purely deterministic generator, denoted \emph{dQRNG}, and a physically coupled, quasi-quantum random number generator, denoted \emph{qQRNG}. This layered model provides a novel physical perspective on how classical computation, randomness extraction, and system-level entropy converge to produce high-quality random numbers.

\subsection{Principle of QPP-RNG}\label{sec:principle}

The core idea behind QPP-RNG models a disordered array $\{ |c_i\rangle \}$ as the ciphertext produced by encrypting an ordered array $\{ |a_i\rangle \}$ with an unknown permutation operator $\hat{p}$:
\begin{equation}
   \{ |c_i\rangle \} = \hat{p} \{ |a_i\rangle \}.
\end{equation}
Permutation sorting thus becomes random decryption: discovering and applying the inverse operator $\hat{p}^{-1}$ to recover the ordered state.

For an array of length $N$, the permutation space $\mathcal{P}$ contains $N!$ elements. Instead of exhaustively searching this space, QPP-RNG uses \emph{random permutation sorting}, which becomes the primary entropy source. Since permutation operators are unitary and reversible, the inverse can be expressed as:
\begin{equation}\label{eq:qpp}
   \hat{p}^{-1} = \prod_{j=1}^{n_p} \hat{p}_j, \quad \hat{p}_j \in \mathcal{P}.
\end{equation}

There are infinitely many such QPP pads, differing in length $n_p$, choice, and order of permutations. Collectively, they span an exponentially large space of decryption paths, introducing significant uncertainty. Convergence to the ordered state requires choosing each $\hat{p}_j$ uniformly at random; biased selection can lead to cycles or stalling. Thus, although QPP-RNG runs entirely in software, it fundamentally depends on a good underlying PRNG: even a simple LCG can suffice, whereas a poor PRNG undermines convergence.

Through this design, QPP-RNG transforms the computationally hard problem of random permutation sorting into a practical entropy harvester, enabling uniform random number generation without external TRNG or QRNG hardware.

\subsection{Definition and Model}

We define the \emph{Quasi-Superposition Quantum-inspired System} (QSQS) as comprising three interacting components:

\begin{itemize}
   \item \textbf{Computing substrate:}  
   The classical computing platform, including CPU, caches, main memory, buses, network interfaces, OS scheduler, and microarchitectural features such as pipelines and execution units. This substrate is subject to real-time stochastic fluctuations from thermal noise, voltage variations, interrupt timing jitter, process migration, and variable memory latency.
   
   \item \textbf{QPP-RNG core:}  
   An algorithmic layer performing random permutation sorting to discover $\hat{p}^{-1}$.
   
   \item \textbf{Physical observables:} Two measurable quantities:
   \begin{enumerate}
      \item \textbf{Modular permutation count per cycle, $\tilde{n}$:}  
      The total permutations $n_p$ used to sort, reduced modulo $2^n$: $\tilde{n} = n_p \bmod 2^n$.
      
      \item \textbf{Modular sorting time per cycle, $\tilde{t}$:}  
      The actual elapsed time (affected by jitter), also reduced modulo $2^n$.
   \end{enumerate}
\end{itemize}

The QSQS state is represented as:
\[
   \ket{\mathrm{QSQS}} = \sum_{i=0}^{2^n-1} \alpha_i \ket{i},
\]
with normalization $\sum_i |\alpha_i|^2=1$, capturing the quasi-superposed possibilities before measurement.

\subsection{System Equation: dQRNG}

We model the process starting from the ordered state $\ket{0,\{a_i\}}$ and the disordered state $\ket{0,\{c_i\}}$. Applying the permutation count operator $\hat{N}_p$ with the inverse pad gives:
\begin{equation}
\begin{aligned}
   \hat{N}_p \hat{p}^{-1} \ket{0,\{c_i\}}
   &= \hat{N}_p \left(\prod_{j=1}^{n_p} \hat{p}_j\right) \ket{0,\{c_i\}} \\
   &= \hat{N}_p \ket{n_p,\{a_i\}} \\
   &= (n_p \bmod 2^n) \ket{n_p \bmod 2^n,\{a_i\}} \\
   &= \tilde{n} \ket{\tilde{n},\{a_i\}}.
\end{aligned}
\end{equation}

All QPP pads yielding the same modular count $\tilde{n}$ belong to the same equivalence class $\widehat{\mathrm{QPP}}_{\tilde{n}}$, forming the eigenbasis:
\[
   \{\ket{0,\{a_i\}}, \ldots, \ket{2^n-1,\{a_i\}}\}.
\]

When the PRNG driving permutation choices is seeded uniformly, the system produces deterministic and reproducible outputs, defining the \textbf{deterministic QRNG (dQRNG)} mode.

\subsection{Uncertainty Principle: qQRNG}

Introducing real-time system jitter means each sorting cycle, while yielding the same $\tilde{n}$, will have different sorting times $\tilde{t}$. This reflects the non-commutation:
\[
   [\hat{N}_p,\hat{T}_p] \neq 0,
\]
analogous to the energy--time uncertainty principle. Thus, $\hat{N}_p$ and $\hat{T}_p$ act as conjugate variables. An eigenstate $\ket{\tilde{n},\{a_i\}}$ of $\hat{N}_p$ can be expressed as:
\[
   \ket{\tilde{n},\{a_i\}}
   = \sum_{j=0}^{2^n-1} \alpha_j \ket{j,\{a_i\}}_t,
\]
where $\ket{j,\{a_i\}}_t$ are eigenstates of the time operator $\hat{T}_p$, and $\alpha_j$ are complex amplitudes.

Crucially, measuring $\hat{T}_p$ when the system is prepared in an eigenstate of $\hat{N}_p$ collapses the quasi-superposed state to a random eigenstate $\ket{j,\{a_i\}}_t$, yielding a non-deterministic output $j$ with probability $\|\alpha_j\|^2$. Empirically, across diverse platforms, these probabilities $\|\alpha_j\|^2$ tend toward uniformity, i.e., $\|\alpha_j\|^2 \approx 1/2^n$, producing outputs that approximate a uniform distribution. This hybrid mechanism---where deterministic structure meets physically induced unpredictability---characterizes the \textbf{quasi-quantum QRNG (qQRNG)} mode.

\subsection{Transition from dQRNG to qQRNG: QPP-RNG}

In practice, QPP-RNG transitions by dynamically reseeding the deterministic generator with entropy from qQRNG:
\begin{itemize}
   \item \emph{dQRNG}: Deterministic output via permutation sorting with a fixed-seed PRNG.
   \item \emph{qQRNG}: Non-deterministic output from measuring sorting times influenced by system jitter.
\end{itemize}

By feeding qQRNG outputs into the PRNG seed, QPP-RNG achieves unpredictable yet statistically uniform randomness:
\[
\mathrm{qQRNG} \xrightarrow{\text{entropy reseeding}} \mathrm{dQRNG}
\longrightarrow \textbf{{qQRNG}}.
\]

This framework explains how classical computation and physical entropy converge to produce platform-independent, system-embedded, software-only high-entropy random number generation.

\section{Experiments and Discussions}
\label{sec:experiments}

To empirically validate the theoretical framework of the Quasi-Superposition Quantum-inspired System (QSQS) and demonstrate the practical viability of QPP-RNG, we implemented a prototype system in Java. We conducted experiments on a representative commodity computing device and systematically evaluated three operational modes: purely deterministic random number generation (dQRNG), purely physical randomness harvesting (qQRNG), and the hybrid approach where dQRNG is dynamically driven by qQRNG-derived entropy.

More comprehensive benchmarking and randomness testing against the NIST SP 800-90B standard have been performed across major computing platforms~\cite{qpp-rng-sci-kuang-2025} and on mobile devices~\cite{kuang2025-qpp-rng-qinp}.

\subsection{Experimental Setup}

Experiments were performed on the following hardware and software platform:

\begin{table}[H]
\centering
\caption{Experimental platform: a Quasi-Superposition Quantum-inspired System (QSQS)}
\label{tab:platform}
\begin{tabular}{@{}lp{4.2cm}@{}}
\toprule
\textbf{Item}   & \textbf{Details} \\ \midrule
Device      & Device-L038-SMS \\
CPU         & Intel i5-1240P (1.7 GHz) \\
RAM         & 16 GB \\
System      & 64-bit, x64-based \\
OS          & Windows 11 Pro, 24H2 \\
OS Build    & 26100.4652 \\
Java Impl.  & QPP-RNG (QSQS Model) \\
\bottomrule
\end{tabular}
\end{table}

This setup represents a typical commercial laptop without specialized hardware accelerators, quantum random number generators (QRNGs), or dedicated true random number generator (TRNG) modules. Within our theoretical framework, the experimental platform is modeled as a \emph{Quasi-Superposition Quantum-inspired System (QSQS)}: a classical computing substrate whose real-time stochastic fluctuations---such as CPU pipeline timing variations, memory latency jitter, dynamic OS scheduling, and cache effects---serve as the physical basis for entropy extraction.

Unlike many TRNG, HRNG, or even QRNG designs that apply whitening, debiasing, or cryptographic post-processing layers to correct imperfections, QPP-RNG directly uses raw physical measurements: the number of random permutations ($n_p$) and the elapsed sorting time. These raw observables are harvested in real time, without subsequent entropy conditioning, ensuring that the final output retains a transparent and verifiable link to the underlying physical noise source. This direct approach not only improves explainability and auditability, but also demonstrates that statistically uniform and unpredictable randomness can emerge purely from system-embedded software tightly coupled to the physical behavior of commodity computing hardware.

\subsection{Algorithmic Flow}
\label{sec:algorithmic-details}

QPP-RNG centers on \emph{random permutation sorting}: starting from a fixed, initially disordered array of small length $N$ (e.g., $N=4,5$), the system repeatedly applies random shuffles until the array becomes sorted. Each complete sorting cycle records two physical observables:
(i) the number of permutations performed ($n_p$) per sorting cycle, and
(ii) the total elapsed real time per sorting cycle.

To produce an $n$-bit random output, this sorting process is repeated $m$ times to accumulate sufficient entropy. The theoretical maximum extractable entropy per output is bounded by $n_{\max} < \log_2 (m \cdot N!)$. For example, generating a 4-bit output typically requires:  $m \ge 4$ with $N=4$, or $m\ge2$ with $N=5$. Larger $N$ or more repetitions $m$ increase entropy at the cost of speed.

The system supports three operational modes:
\begin{itemize}
   \item \emph{dQRNG}: deterministic; uses a PRNG seeded with a fixed secret; outputs the modular permutation count $\tilde{n}$.
   \item \emph{qQRNG}: purely physical; outputs the modular elapsed time $\tilde{t}$, capturing live system jitter.
   \item \emph{QPP-RNG (integrated)}: hybrid; dynamically reseeds the PRNG with $\tilde{t}$, blending physical entropy with deterministic uniformity.
\end{itemize}

Algorithm~\ref{alg:getQppRNG} summarizes the core logic:

\vspace{0.5em}

\begin{algorithm}[H]
\caption{\texttt{getQppRNG}: Generate $n$-bit random outputs via permutation sorting}
\label{alg:getQppRNG}
\begin{algorithmic}[1]
\Require Disordered array $A$, number of outputs $num$, repetitions $m$, bit length $n$
\Ensure Array $randN$ of $n$-bit random outputs
\State Initialize $\mathit{seed}$ (fixed for dQRNG/qQRNG; updated in QPP-RNG)
\For{$i=1$ to $num$}
   \State $start \leftarrow$ current time
   \State $n_p \leftarrow 0$
   \For{$k=1$ to $m$}
     \State $array \leftarrow$ copy of $A$
     \While{not sorted}
       \State \texttt{permutArray}($array$)
       \State $n_p \leftarrow n_p+1$
     \EndWhile
   \EndFor
   \State $\tilde{n} \leftarrow n_p \bmod 2^n$
   \State $end \leftarrow$ current time
   \State $\tilde{t} \leftarrow (end - start) \bmod 2^n$
   \If{QPP-RNG}
     \State $\mathit{seed} \leftarrow (\mathit{seed} \ll n) \oplus \tilde{t}$
   \EndIf
   \If{qQRNG}
     \State $randN[i] \leftarrow \tilde{t}$
   \Else \Comment{dQRNG or QPP-RNG}
     \State $randN[i] \leftarrow \tilde{n}$
   \EndIf
\EndFor
\end{algorithmic}
\end{algorithm}

\begin{algorithm}[H]
\caption{\texttt{permutArray}: Fisher--Yates shuffle driven by PRNG}
\label{alg:permutArray}
\begin{algorithmic}[1]
\Require Array $arr$ of length $N$
\Ensure Randomly permuted $arr$
\For{$j=N-1$ down to $1$}
   \State $r \leftarrow$ \texttt{nextInt}($0, j$) \Comment{Random index from PRNG}
   \State Swap $arr[j]$ and $arr[r]$
\EndFor
\end{algorithmic}
\end{algorithm}

\begin{algorithm}[H]
\caption{\texttt{nextInt}: Draw random integer in $[0, bound]$ from LCG PRNG}
\label{alg:nextInt}
\begin{algorithmic}[1]
\Require Bound $bound$
\Ensure Integer in $[0, bound]$
\State $\mathit{seed} \leftarrow (a \times \mathit{seed} + c) \bmod m$
\State \Return $\mathit{seed} \bmod (bound+1)$
\end{algorithmic}
\end{algorithm}

Here, \texttt{permutArray} applies a Fisher--Yates shuffle using a seeded PRNG; \texttt{nextInt} draws indices via an LCG:
\[
\mathit{seed} \leftarrow (a \times \mathit{seed} + c) \bmod m.
\]
The parameters $N$ and $m$ balance entropy and speed; the three modes allow purely deterministic, purely physical, or hybrid entropy generation---all in system-embedded software.

\subsection{From Right-skewed Distributions to Uniform Distributions}
\label{sec:distribution}

In the QSQS system, each random number is generated by sorting a chosen initially disordered array of small size $N$ (e.g., typically $N=4$ elements in this experiment) via random permutation sorting.
Importantly, the specific degree of disorder in the initial array does not affect the overall statistical behaviour; it suffices that the array is not already sorted.

From the \emph{same} sorting cycles, we collect two raw physical measurements:
\begin{enumerate}
   \item the number of random permutations required to complete sorting, denoted $\hat{N}_p$, and
   \item the elapsed time, $\hat{T}_i$, measured by a high-resolution system timer.
\end{enumerate}

For this study, we primarily used a 4-element array to produce 4-bit outputs. In earlier benchmarking work, we also used larger arrays (e.g., $N=5$) and more repetitions (e.g., $m=5$) to generate 8-bit outputs~\cite{qpp-rng-sci-kuang-2025}.

To increase unpredictability, each sorting cycle is repeated $m=4$ times, enlarging the random search space from $4!=24$ to $4! \times 4 = 96$. Instead of treating these as four independent outputs, they are aggregated into a single cycle, yielding an entropy estimate of approximately $\log_2 96 \approx 6.5$ bits. Meanwhile, repeated sorting accumulates fresh system jitter, further improving unpredictability.

\begin{figure}[ht]
   \centering
   \includegraphics[scale=0.3]{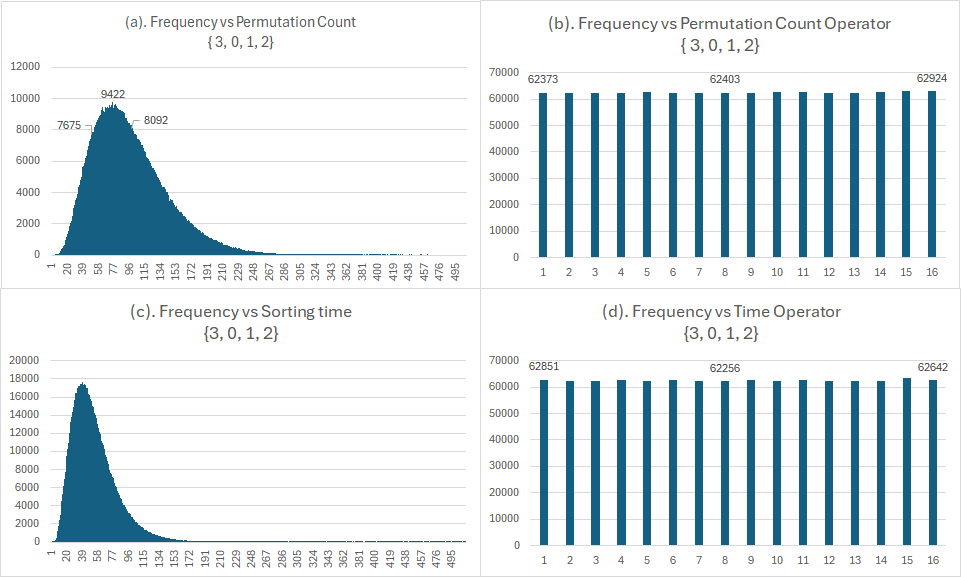}
   \caption{Raw distributions of random permutation counts and elapsed times per sorting cycle, with corresponding modulo 16 reductions. Panels (a) and (c): right-skewed distributions of raw permutation counts $\hat{N}_p$ and elapsed times $\hat{T}_i$. Panels (b) and (d): nearly uniform distributions after modulo 16, producing 4-bit random outputs.}
   \label{fig:raw_distributions}
\end{figure}

Figure~\ref{fig:raw_distributions} shows the raw distributions: permutation counts $\hat{N}_p$ exhibit a right-skewed shape, with many fast sortings and a long tail from cycles requiring more permutations (Panel~(a)). This resembles waiting-time processes (e.g., geometric or negative binomial distributions). The elapsed time distribution $\hat{T}_i$ is similarly right-skewed (Panel~(c)): longer sorting cycles naturally take more time, while fine-grained system-level jitter introduces true physical entropy.

Crucially, after applying modulo 16, the observable outputs become nearly uniform (Panels~(b) and~(d)). For the deterministic branch, the observable output $\tilde{n} = n_p \bmod 16$ emerges from many possible QSQS states \(\lvert n_p,\{a_i\} \rangle\) whose permutation counts differ by multiples of 16. These degenerate states—distinct internal configurations that produce the same observable $\tilde{n}$—collectively “fill out” the output space and flatten the initially skewed distribution into a uniform one. This is an elegant effect of the QSQS structure: the system’s internal degeneracies transform biased raw counts into uniform observable outputs.

For the non-deterministic branch, the elapsed time observable $\hat{T}_i \bmod 16$ shows a different but equally remarkable effect. Although the raw elapsed times $\hat{T}_i$ are also right-skewed—shaped by both algorithmic properties and unpredictable system jitter—the modulo operation, together with live physical fluctuations, scatters these measurements across the 4-bit output space. The fine-grained, non-repeatable noise effectively “fills the gaps,” helping the modulo output escape the skewness seen in raw counts. 

What makes this especially striking is that the two skewed distributions—one from deterministic algorithmic counts, and one from physical timing measurements—are visually and statistically different in the raw data (Panels~(a) and~(c)). Yet both converge to near-uniform distributions after the same simple modulo transformation (Panels~(b) and~(d)). This near-miraculous uniformity emerges from a combination of structural degeneracy (for $\tilde{n}$) and physical entropy injection (for $\hat{T}_i$). 

Together, these dual observables—deterministic $\tilde{n}$ from permutation counts and non-deterministic $\hat{T}_i$ from physical jitter—demonstrate the QSQS model’s synergy between mathematical permutation randomness and real-world entropy. This duality enables QPP-RNG to produce high-quality, uniform random numbers, rooted both in algorithmic structure and in live physical noise—like a software-based “miracle” of randomness without needing specialized hardware.

\subsection{Deterministic vs Non-deterministic Randomness}\label{sec:det-non-det}

Table~\ref{tab:det-non-det} presents an empirical illustration of the Quasi-Superposition Quantum-inspired System (QSQS) model, highlighting its core mechanism: the coexistence of deterministic and non-deterministic randomness derived from the same sorting cycles.

In this example, the QSQS system repeatedly sorts the same disordered $4$-element array \(\{3,0,1,2\}\), applying random permutation sorting with \(m=4\) repetitions per sorting cycle. Each row in Table~\ref{tab:det-non-det} corresponds to one such sorting cycle, labeled \(\widehat{\mathrm{QPP}}_{1}\) to \(\widehat{\mathrm{QPP}}_{10}\), each initialized with a different QPP pad (i.e., a distinct sequence of permutations). For each QPP pad, we repeated the sorting cycle 10 times to observe both deterministic and non-deterministic behaviors.

\begin{table}[htbp]
\centering
\caption{A QSQS system with a disordered array \(\{3,0,1,2\}\), sorting with \(m=4\) repetitions to produce \(4\)-bit deterministic random numbers with \(\hat{N}_p\) and non-deterministic random numbers with \(\hat{T}_i: i \in [1,10]\).}
\label{tab:det-non-det}
\renewcommand{\arraystretch}{1.4}
\begin{tabular}{|c|c|c|c|c|c|c|c|c|c|c|c|}
\hline
    & \(\hat{N}_p\) & \(\hat{T}_1\) & \(\hat{T}_2\) & \(\hat{T}_3\) & \(\hat{T}_4\) & \(\hat{T}_5\) & \(\hat{T}_6\) & \(\hat{T}_7\) & \(\hat{T}_8\) & \(\hat{T}_9\) & \(\hat{T}_{10}\) \\
\hline
$\widehat{\mathrm{QPP}}_{1}$ & 9  & 7 & 0 & 3 & 2 & 1 & 8 & 4 & 7 & 12 & 8 \\
$\widehat{\mathrm{QPP}}_{2}$ & 1  & 12 & 15 & 1 & 7 & 8 & 4 & 2 & 11 & 0 & 7 \\
$\widehat{\mathrm{QPP}}_{3}$ & 14 & 14 & 8 & 2 & 1 & 5 & 8 & 15 & 8 & 2 & 12 \\
$\widehat{\mathrm{QPP}}_{4}$ & 9  & 14 & 10 & 2 & 11 & 13 & 10 & 5 & 15 & 13 & 3 \\
$\widehat{\mathrm{QPP}}_{5}$ & 11 & 13 & 15 & 12 & 1 & 0 & 0 & 12 & 9 & 11 & 3 \\
$\widehat{\mathrm{QPP}}_{6}$ & 13 & 0 & 12 & 0 & 0 & 5 & 6 & 2 & 15 & 7 & 8 \\
$\widehat{\mathrm{QPP}}_{7}$ & 7  & 13 & 4 & 0 & 2 & 11 & 11 & 11 & 4 & 5 & 14 \\
$\widehat{\mathrm{QPP}}_{8}$ & 10 & 13 & 2 & 8 & 5 & 2 & 13 & 8 & 13 & 12 & 10 \\
$\widehat{\mathrm{QPP}}_{9}$ & 11 & 14 & 12 & 4 & 15 & 13 & 11 & 6 & 0 & 13 & 4 \\
$\widehat{\mathrm{QPP}}_{10}$ & 4  & 4 & 3 & 3 & 11 & 10 & 9 & 12 & 13 & 14 & 2 \\
\hline
\end{tabular}
\end{table}

The first column shows the deterministic random number $\hat{N}_p$, obtained from the total permutation count modulo 16. Importantly, this output remains identical across all repetitions of the same QPP pad, reflecting the deterministic, algorithmic randomness imposed by the sorting logic and the fixed pad. This is why it is referred to as a \emph{deterministic QRNG} (dQRNG).

Although each QPP pad deterministically brings the QSQS system to a state \(\lvert n_p, \{a_i\} \rangle\), the observable output is \(\tilde{n} = n_p \bmod 16\). Different internal states with distinct permutation counts \(n_p\) (e.g., \(9, 25, 41,\ldots\)) can produce the same observable \(\tilde{n} = 9\). These are not collapsed states, but rather \emph{degenerate states} — different eigenstates of the QSQS system that share the same eigenvalue under the modulo-16 observable. For example, $\widehat{\mathrm{QPP}}_{1}$ and $\widehat{\mathrm{QPP}}_{4}$ both produce the output 9, while $\widehat{\mathrm{QPP}}_{5}$ and $\widehat{\mathrm{QPP}}_{9}$ both produce 11, despite arising from different QPP pads and potentially very different internal permutation paths.

This degeneracy plays a remarkable role: although the underlying distribution of \(n_p\) may be skewed due to structural or algorithmic biases, taking the output modulo 16 leverages these degenerate states to transform the distribution into a more uniform one over \(\tilde{n}\). In effect, the many degenerate states collectively fill the gaps and smooth out the distribution of observable outputs — an emergent phenomenon central to achieving uniform randomness from deterministic processes.

In contrast, the remaining columns display the elapsed time measurements \(\hat{T}_i\) (also modulo 16) for ten independent repetitions per QPP pad. These outputs vary significantly from run to run, even under identical algorithmic conditions, because they capture unpredictable physical noise and microarchitectural fluctuations, such as CPU pipeline jitter, cache effects, and branch prediction. This non-deterministic randomness is why the elapsed time-based output is called a \emph{quantum-like or quasi QRNG} (qQRNG).

Together, these observations capture the dual nature of the QSQS system: a deterministic algorithmic branch (via $\hat{N}_p$) and a non-deterministic physical branch (via $\hat{T}$). This duality enables QPP-RNG to produce high-quality random numbers rooted both in mathematical permutation randomness and in physical entropy, aligning with the broader goal of system-embedded, software-based true random number generation without dedicated hardware.

\subsection{Entropy Convergence for dQRNG and qQRNG}\label{sec:converge}

Within the Quasi-Superposition Quantum-inspired System (QSQS) framework, random numbers are generated by measuring two distinct observables from the same permutation sorting process (see Section~\ref{sec:det-non-det} which defined two operational modes:

\textbf{dQRNG mode.}
In this mode, random numbers are produced deterministically by applying random permutation sorting under the control of a Quantum Permutation Pad (QPP). The sequence of permutations \(\{\hat{p}_j\}\) is generated by a pseudorandom number generator (PRNG) initialized with a fixed secret seed.
\begin{itemize}
   \item The output is strictly reproducible: rerunning the generator with the same seed produces the same sequence of permutation counts modulo \(2^n\).
   \item Empirical tests show that these deterministic outputs exhibit strong uniformity across the \(2^n\) possible states, thanks to uniformly distributed permutation selections.
\end{itemize}

\textbf{qQRNG mode.}
Here, randomness is harvested from physical system-level jitter by measuring the elapsed sorting time \(t\) modulo \(2^n\). Since real-world execution times vary unpredictably due to CPU pipeline fluctuations, memory latency, cache contention, and OS scheduling effects, these outputs are fundamentally non-deterministic and independent of the secret seed.

Table~\ref{tab:dqrng-qqrng} presents empirical results demonstrating how entropy converges in both modes as the repetition factor \(m\) increases from 1 to 6, with a fixed array size \(N=4\). Each sorting cycle thereby explores a search space enlarged from \(N!=24\) to \(m \cdot N!\). The theoretical entropy bound, computed as \(2\log_2(m \cdot N!)\), correspondingly increases for 8-bit outputs derived from these 4-bit entropy sources.

To evaluate convergence toward uniform distributions, we selected three highly sensitive statistical measures commonly used in random number testing:
\begin{itemize}
   \item Shannon entropy --- capturing the average unpredictability of symbols;
   \item NIST SP 800-90B estimated min-entropy --- reflecting worst-case predictability;
   \item Chi-squared (\(\chi^2\)) statistic --- measuring deviation from an ideal uniform distribution.
\end{itemize}

As shown in the table:
\begin{itemize}
   \item Both dQRNG and qQRNG modes show steady increases in Shannon and min-entropy as \(m\) increases, approaching the theoretical bound.
   \item The \(\chi^2\) statistic drops sharply, indicating that the empirical distributions become increasingly uniform.
\end{itemize}

Notably, despite the fundamental difference in entropy source — deterministic algorithmic randomness in dQRNG mode versus physical non-deterministic randomness in qQRNG mode — the convergence trends remain similar once \(m \geq 4\). This convergence highlights the effect of enlarging the random search space through repeated sortings, which both increases permutation complexity and accumulates additional system jitter.

\begin{figure}[ht]
   \centering
   \includegraphics[scale=0.25]{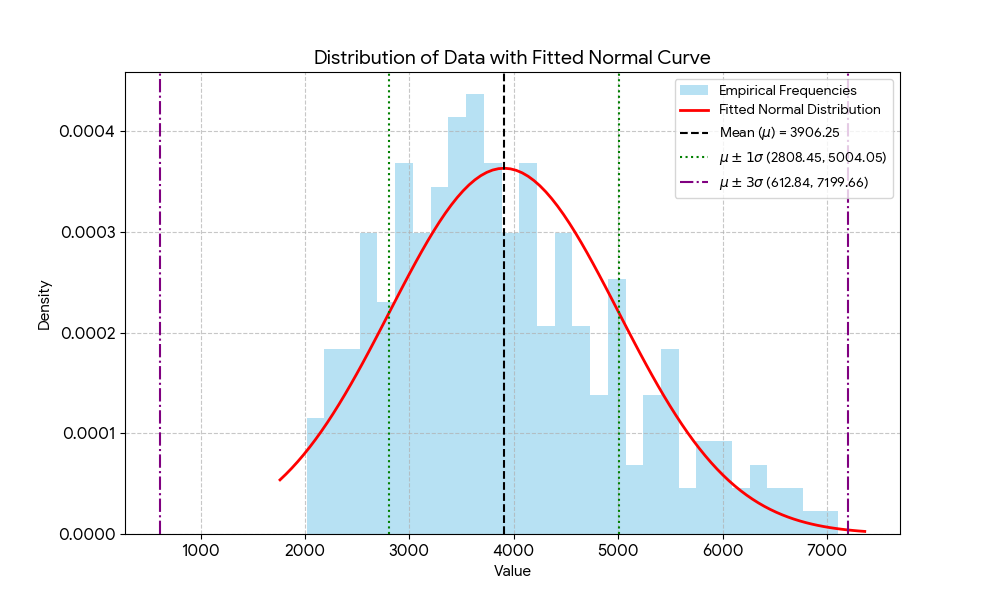}
   \caption{Byte-level frequency distribution from dQRNG with \(N=4, m=1\). The wide spread (\(\sigma=1097.80\)) indicates significant deviation from uniformity.}
   \label{fig:dQRNG-41}
\end{figure}

\begin{figure}[ht]
   \centering
   \includegraphics[scale=0.25]{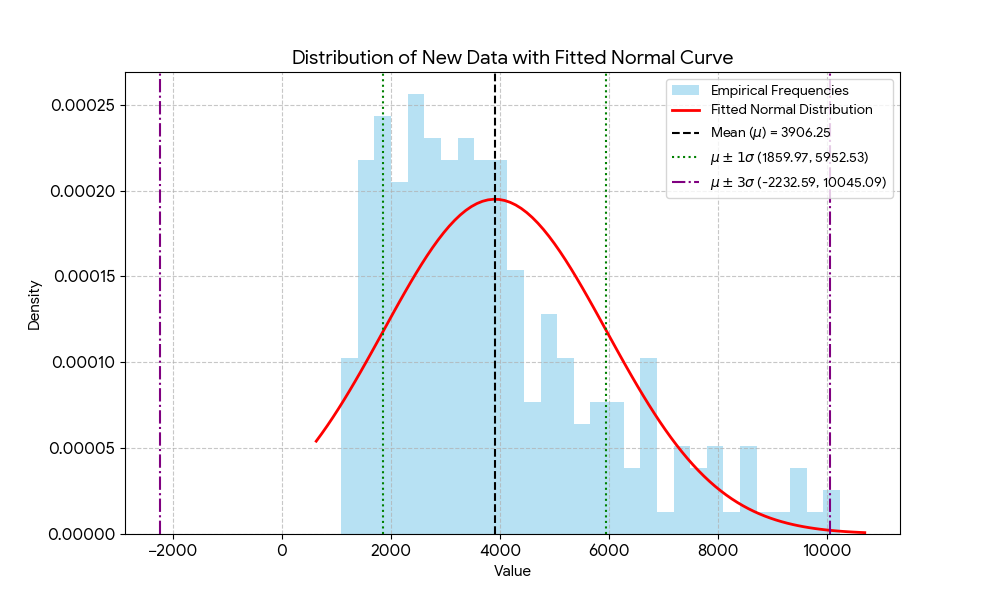}
   \caption{Byte-level frequency distribution from qQRNG with \(N=4, m=1\). Even larger spread (\(\sigma=2046.28\)) shows higher initial non-uniformity due to system-level jitter.}
   \label{fig:qQRNG-41}
\end{figure}

\begin{figure}[ht]
   \centering
   \includegraphics[scale=0.25]{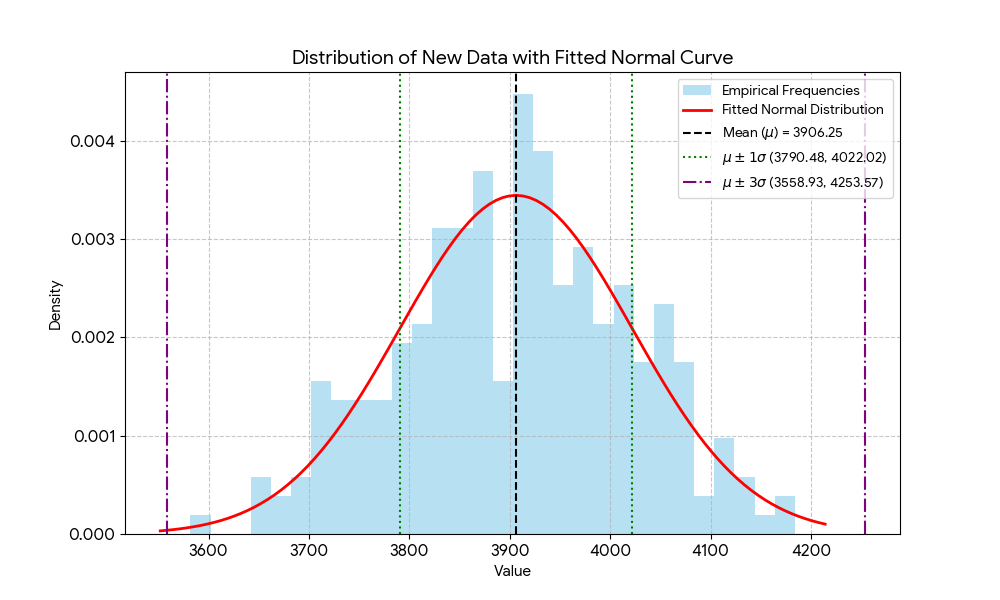}
   \caption{For \(N=4, m=2\), dQRNG frequencies become substantially flatter (\(\sigma=115.77\)), indicating rapid convergence toward uniform distribution.}
   \label{fig:dQRNG-42}
\end{figure}

\begin{figure}[ht]
   \centering
   \includegraphics[scale=0.25]{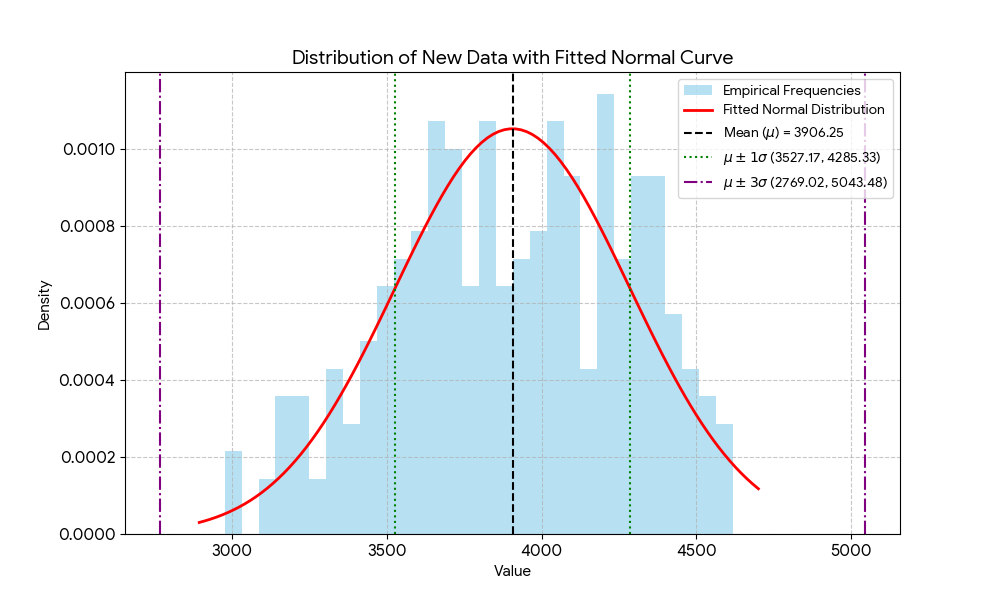}
   \caption{For \(N=4, m=2\), qQRNG shows notable improvement (\(\sigma=379.08\)), though still higher variance than dQRNG.}
   \label{fig:qQRNG-42}
\end{figure}

\begin{figure}[ht]
   \centering
   \includegraphics[scale=0.25]{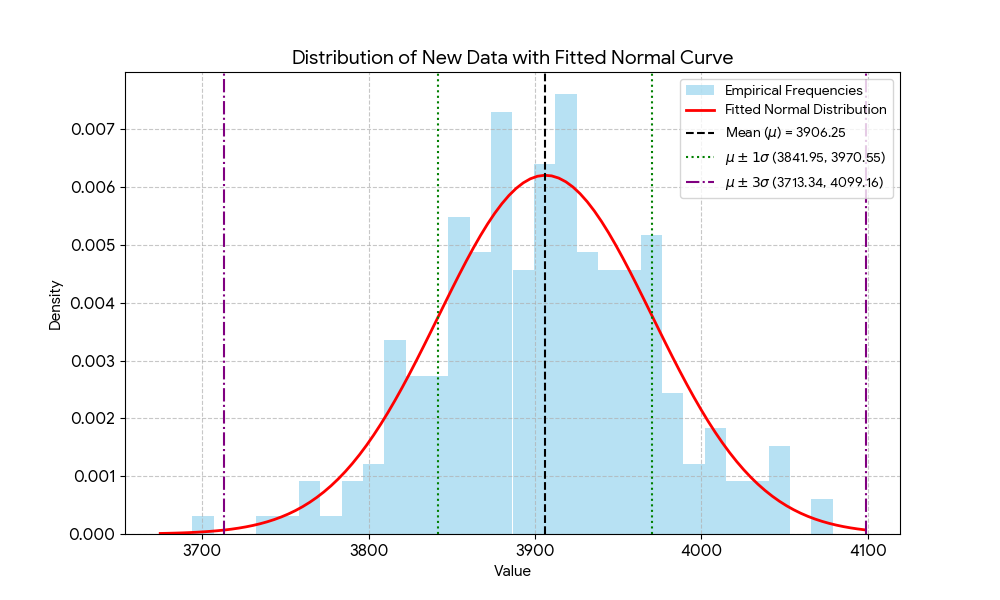}
   \caption{At \(N=4, m=3\), dQRNG achieves near-uniform distribution (\(\sigma=64.30\)), very close to the theoretical ideal.}
   \label{fig:dQRNG-43}
\end{figure}

\begin{figure}[ht]
   \centering
   \includegraphics[scale=0.25]{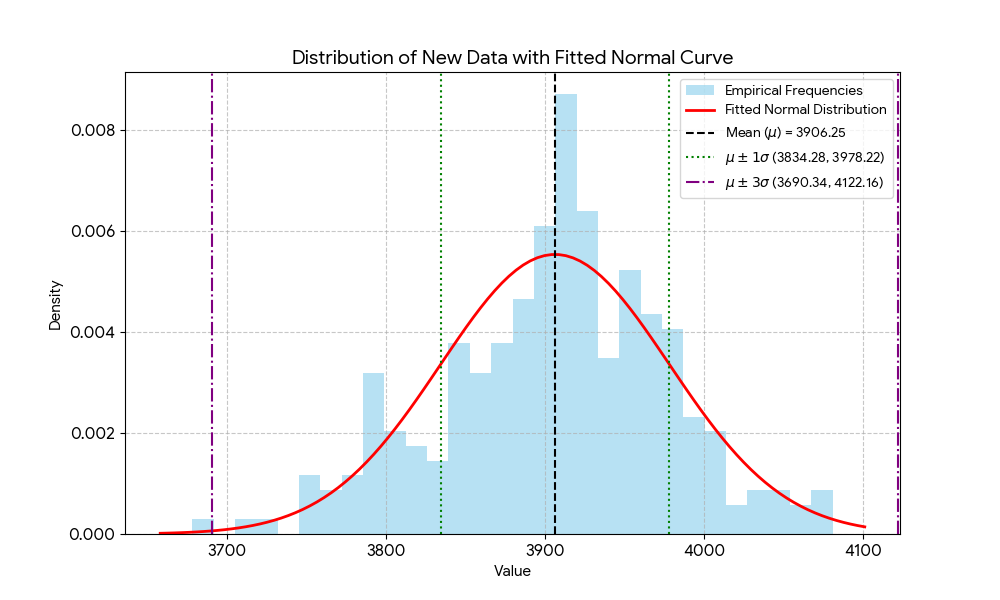}
   \caption{At \(N=4, m=3\), qQRNG variance further drops (\(\sigma=116.97\)), though still above the ideal uniformity.}
   \label{fig:qQRNG-43}
\end{figure}

\begin{figure}[ht]
   \centering
   \includegraphics[scale=0.25]{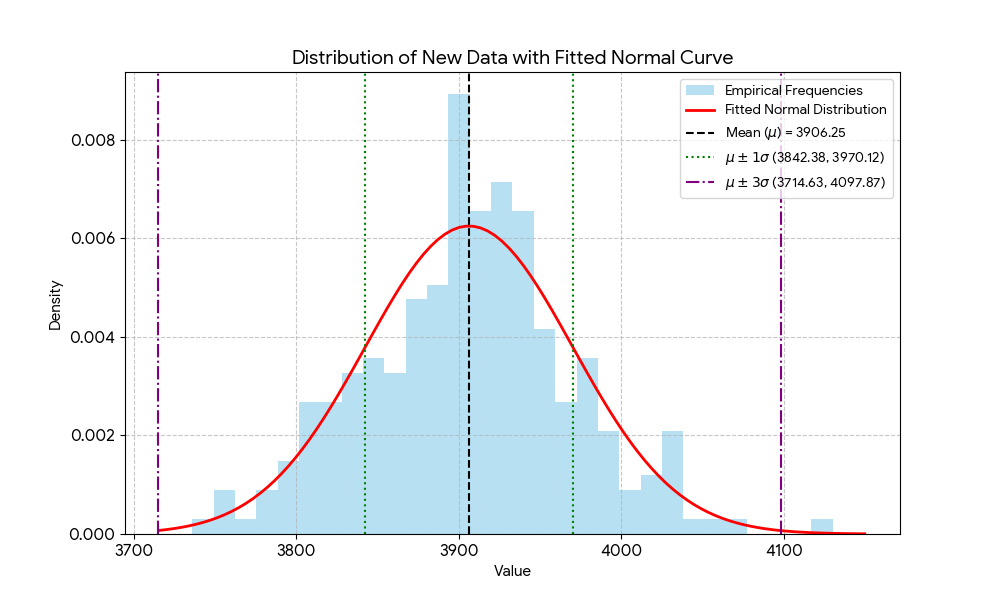}
   \caption{For \(N=4, m=4\), dQRNG frequency distribution stabilizes (\(\sigma=63.87\)), matching the theoretical uniform distribution.}
   \label{fig:dQRNG-44}
\end{figure}

\begin{figure}[ht]
   \centering
   \includegraphics[scale=0.25]{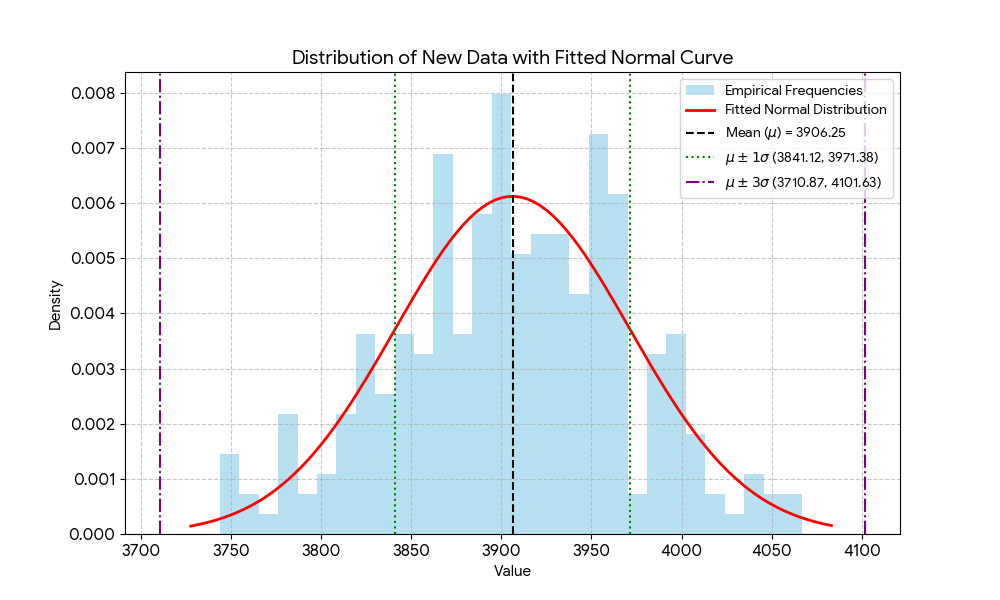}
   \caption{For \(N=4, m=4\), qQRNG also converges closely (\(\sigma=65.13\)), effectively matching the uniform distribution’s theoretical standard deviation.}
   \label{fig:qQRNG-44}
\end{figure}

\begin{table*}[t]
\centering
\caption{Entropy convergence as the repetition \(m\) increases with \(N=4\). The theoretical bound \(2\log_2(m \cdot N!)\) is compared against measured Shannon entropy, NIST SP 800-90B min-entropy, and chi-squared statistics for both dQRNG and qQRNG modes.}
\label{tab:dqrng-qqrng}
\renewcommand{\arraystretch}{1.1}
\begin{tabular}{|l|l|c|c|c|}
\hline
    & \((N, m)\) & \(2\log_2(m \cdot N!)\) & dQRNG & qQRNG \\
\hline
Shannon Entropy & (4, 1) & 9.170 & 7.944563 & 7.818526 \\
90B min-Entropy & (4, 1) &   & 7.136340 & 6.608373 \\
\(\chi^2\)    & (4, 1) &   & 78678.58 & 270079.82 \\
\hline
Shannon Entropy & (4, 2) & 11.170 & 7.999368 & 7.993158 \\
90B min-Entropy & (4, 2) &    & 7.900901 & 7.758516 \\
\(\chi^2\)    & (4, 2) &    & 875.01  & 9381.27 \\
\hline
Shannon Entropy & (4, 3) & 12.398 & 7.999805 & 7.999744 \\
90B min-Entropy & (4, 3) &    & 7.937568 & 7.938630 \\
\(\chi^2\)    & (4, 3) &    & 269.94  & 353.86 \\
\hline
Shannon Entropy & (4, 4) & 13.170 & 7.999800 & 7.999796 \\
90B min-Entropy & (4, 4) &    & 7.919642 & 7.926294 \\
\(\chi^2\)    & (4, 4) &    & 266.35  & 282.67 \\
\hline
Shannon Entropy & (4, 5) & 13.8148 & 7.999807 & 7.999799 \\
90B min-Entropy & (4, 5) &    & 7.933683 & 7.917548 \\
\(\chi^2\)    & (4, 5) &    & 266.55  & 278.73 \\
\hline
Shannon Entropy & (4, 6) & 14.398 & 7.999814 & 7.999825 \\
90B min-Entropy & (4, 6) &    & 7.922789 & 7.933683 \\
\(\chi^2\)    & (4, 6) &    & 257.33  & 242.09 \\
\hline
\end{tabular}
\end{table*}
To further illustrate this convergence visually, Figures~\ref{fig:dQRNG-41}–\ref{fig:qQRNG-44} present bell curve plots of byte-level frequency distributions for both dQRNG and qQRNG modes as the repetition factor \(m\) increases. Each plot reports the standard deviation \(\sigma\) of byte frequencies, where the theoretical ideal for a perfectly uniform distribution over a 1~MB sample (approximately one million bytes distributed across 256 bins) is about \(\sigma \approx 63.6\). Observing how the empirical \(\sigma\) values decrease and converge toward this theoretical bound provides an intuitive view of the entropy convergence process.

Figures~\ref{fig:dQRNG-41} and~\ref{fig:qQRNG-41} show the initial distributions for \(N=4, m=1\), where the standard deviations are 1097.80 for dQRNG and 2046.28 for qQRNG—both substantially larger than the ideal value. This highlights strong initial non-uniformity and aligns with the lower measured Shannon entropy, min-entropy, and high \(\chi^2\) statistics reported earlier.

When \(m\) increases to 2 (Figures~\ref{fig:dQRNG-42} and~\ref{fig:qQRNG-42}), the distributions become visibly flatter: \(\sigma\) drops to 115.77 for dQRNG and 379.08 for qQRNG, indicating significant improvement, though still above the ideal. At \(m=3\) (Figures~\ref{fig:dQRNG-43} and~\ref{fig:qQRNG-43}), dQRNG achieves near-uniformity with \(\sigma=64.30\), very close to the theoretical target, while qQRNG also improves markedly with \(\sigma=116.97\).

Finally, for \(m=4\) (Figures~\ref{fig:dQRNG-44} and~\ref{fig:qQRNG-44}), both modes effectively converge: dQRNG reaches \(\sigma=63.87\) and qQRNG reaches \(\sigma=65.13\), closely matching the theoretical uniform distribution. Beyond \(m=4\), further increases bring negligible additional improvement, confirming that the entropy has converged.

\medskip

These visualizations confirm both quantitatively and intuitively that repeated permutation sorting (larger \(m\)) drives dQRNG and qQRNG outputs toward ideal uniformity. This trend, which we refer to as QSQS entropy convergence, complements the statistical results summarized in Table~\ref{tab:dqrng-qqrng}.

\vspace{1ex}

\subsection{QPP-RNG: System-Embedded TRNG}

Building on the QSQS framework, we construct \emph{QPP-RNG} by combining the strengths of both dQRNG and qQRNG modes. Specifically, qQRNG's physically non-deterministic outputs are used to reseed the PRNG driving dQRNG before each sorting cycle. This transforms the originally deterministic dQRNG into a non-deterministic generator: each sorting cycle starts from a fresh, unpredictable seed derived from real-time system jitter, while still benefiting from the uniformity of algorithmic random permutation sorting. This synergy produces a hybrid true random number generator, entirely in software, which we term \emph{QPP-RNG}.

\medskip

Table~\ref{tab:qpp-rn} summarizes the statistical results of QPP-RNG under various configurations. We evaluate two array sizes, \(N=4\) and \(N=5\), each with repetition factors \(m=3,4,5\). For \(N=4\), the entropy source per sorting cycle remains 4 bits; for \(N=5\), it becomes 8 bits, allowing each cycle to directly generate an 8-bit random output.

Across all configurations, QPP-RNG demonstrates strong entropy convergence:
\begin{itemize}
   \item The measured Shannon entropy approaches the ideal value of 8 bits, indicating near-maximal unpredictability.
   \item The NIST SP 800-90B estimated min-entropy remains high, reflecting low worst-case predictability.
   \item The chi-squared (\(\chi^2\)) statistic steadily decreases with larger \(m\), showing that distributions become increasingly uniform.
\end{itemize}

As \(N\) and \(m\) increase, the search space expands exponentially (e.g., from \(m \cdot 4! = m \cdot 24\) for \(N=4\) to \(m \cdot 120\) for \(N=5\)), improving entropy convergence but reducing generation speed due to higher computational complexity. For instance, with \(N=4, m=3\), QPP-RNG achieves 128~KB/s, whereas with \(N=5, m=5\), speed drops to 16~KB/s.

\medskip

\begin{figure}[ht]
   \centering
   \includegraphics[scale=0.25]{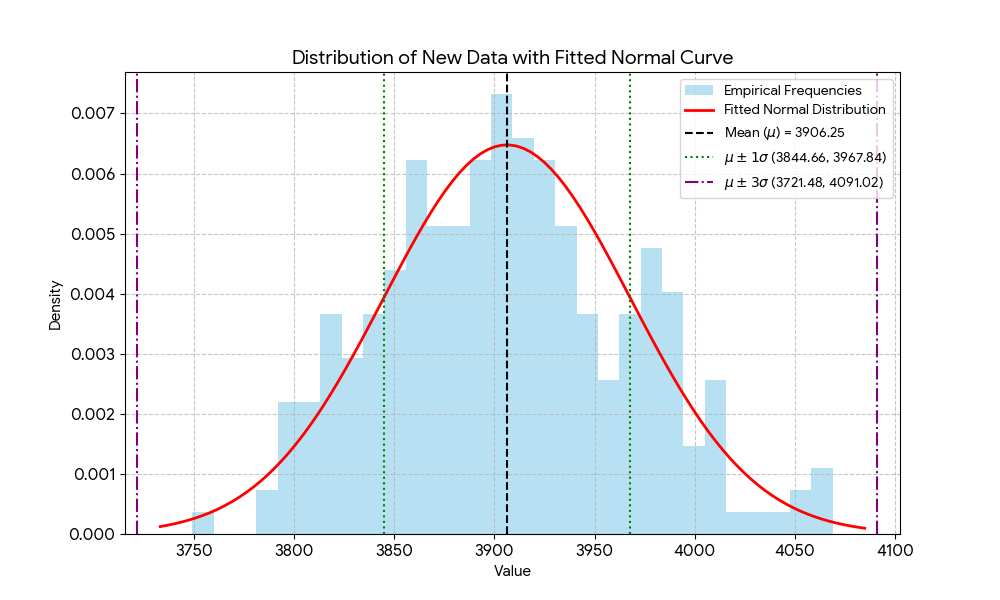}
   \caption{Byte-level frequency distribution from QPP-RNG with \(N=4, m=3\). The narrow spread (\(\sigma=61.59\)) indicates convergence to near-ideal uniformity.}
   \label{fig:qpp-rng-43}
\end{figure}

\begin{figure}[ht]
   \centering
   \includegraphics[scale=0.25]{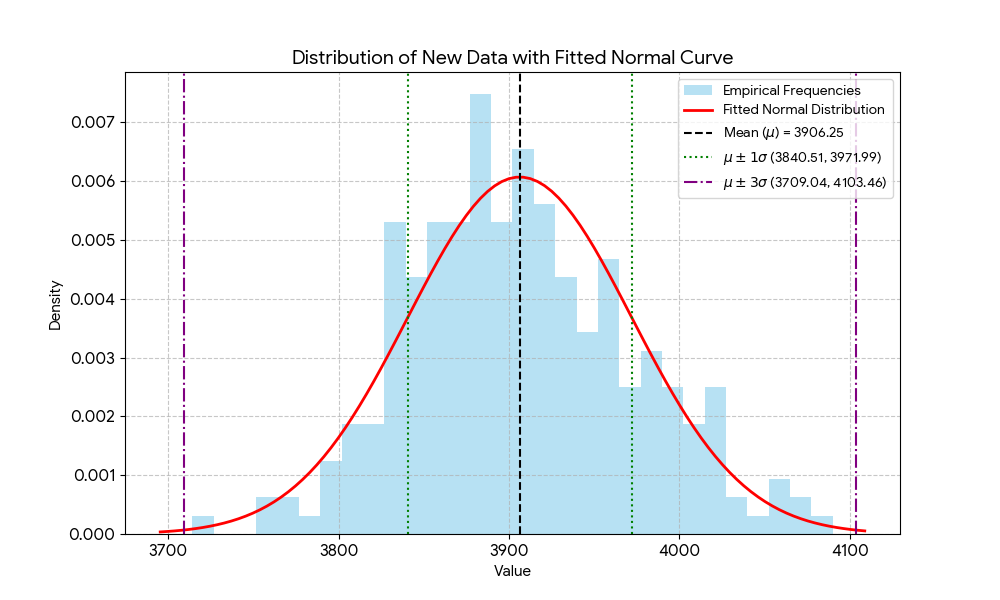}
   \caption{Byte-level frequency distribution from QPP-RNG with \(N=4, m=4\). The standard deviation remains low (\(\sigma=65.47\)), confirming stable uniformity.}
   \label{fig:qpp-rng-44}
\end{figure}

Figures~\ref{fig:qpp-rng-43} and \ref{fig:qpp-rng-44} visualize the convergence by showing bell curve distributions of byte-level frequencies. For \(N=4, m=3\) in Figure~\ref{fig:qpp-rng-43}, the standard deviation \(\sigma=61.59\) is already below the theoretical ideal for a perfectly uniform distribution (\(\approx 63.6\)). For \(m=4\) in Figure~\ref{fig:qpp-rng-44}, \(\sigma\) remains close (\(65.47\)), confirming stability and convergence.

\medskip

In this hybrid mode, non-deterministic outputs from qQRNG (specifically, $sysJitter$) are used to reseed the PRNG driving dQRNG:
\begin{itemize}
   \item This preserves the strong uniformity and structure of the deterministic sorting process.
   \item At the same time, it introduces physical unpredictability, making the output practically irreproducible.
   \item Empirical results show high entropy, uniform distributions, and non-reproducibility across runs with live reseeding.
\end{itemize}

Together, these results demonstrate that QPP-RNG effectively merges deterministic and non-deterministic entropy sources into a robust software-based TRNG, suitable for cryptographic and high-security applications.

\begin{table*}[t]
\centering
\caption{QPP-RNG entropy convergence demonstration as the repetition \(m\) increases. For \(N=4\), the entropy source per cycle is 4 bits; for \(N=5\), it is 8 bits, producing 8-bit outputs directly.}
\label{tab:qpp-rn}
\renewcommand{\arraystretch}{1.1}
\begin{tabular}{|l|c|c|c|c|}
\hline
\((N, m)\) & \textbf{Shannon Entropy} & \textbf{90B min-Entropy} & \textbf{\(\chi^2\)} & \textbf{Speed (KB/s)} \\
\hline
(4, 3) & 7.9998215 & 7.9411100 & 247.65 & 128 \\
(4, 4) & 7.9997967 & 7.9336834 & 282.12 & 94 \\
(4, 5) & 7.9998197 & 7.9464382 & 249.84 & 58 \\
(5, 3) & 7.9982540 & 7.8796799 & 2407 & 30 \\
(5, 4) & 7.9996224 & 7.9002119 & 523.95 & 20 \\
(5, 5) & 7.9998058 & 7.9432389 & 268.88 & 16 \\
\hline
\end{tabular}
\end{table*}

\medskip

In summary, QPP-RNG demonstrates how the QSQS design can unify deterministic algorithmic structures with physical non-determinism to produce a high-quality, software-based true random number generator. By reseeding the deterministic permutation process (dQRNG) with live, unpredictable system jitter harvested from qQRNG, QPP-RNG maintains strong uniformity, achieves high entropy, and ensures practical unpredictability. This hybrid approach not only bridges the gap between deterministic and non-deterministic randomness, but also lays a scalable foundation for cryptographic applications requiring reproducible statistical quality, resilience to bias, and true entropy without specialized hardware. Subsequent sections will describe the overall architecture, implementation strategies, and performance benchmarks of QPP-RNG in practical systems.

\section{Conclusion}\label{sec:conclusion}

In this work, we introduced and experimentally validated the \emph{Quasi-Superposition Quantum-inspired System (QSQS)} --- a conceptual quantum system in which randomness emerges from measuring two conjugate observables of the same physical-computational process: the deterministic permutation count \(n_p\) and the fundamentally non-deterministic sorting time \(t\). Analogous to quantum systems, these observables obey an uncertainty-like constraint: the more precisely one is determined, the more the other can vary, making the joint system a sustainable source of entropy.

\medskip

Importantly, we showed that \emph{QPP-RNG} not only draws inspiration from QSQS but concretely \emph{realizes} it as a practical, system-embedded, software-based true random number generator (TRNG). In this design, real-time measurements of sorting time \(t\) --- shaped by microarchitectural factors such as CPU pipeline jitter and OS scheduling --- dynamically reseed the PRNG that deterministically drives the permutation sequence. This hybrid structure harnesses the interplay of algorithmic determinism and live physical entropy: randomness emerges organically from the quasi-superposition structure of the system.

\medskip

Crucially, we highlighted how QSQS turns raw right-skewed distributions --- from permutation counts and elapsed times --- into nearly uniform observable outputs after modulo reduction. This effect arises from the system's internal degeneracies: many distinct internal states map to the same observable output, collectively “filling out” the output space and flattening biases. This transformation from biased raw measurements to uniform distributions is a central feature of QSQS, showing how structural degeneracy and physical entropy injection together produce high-quality randomness.

\medskip

Empirical evaluation demonstrated that as the repetition factor \(m\) increases, the output entropy of QPP-RNG converges toward theoretical maxima: Shannon and min-entropy values approach 8 bits, chi-squared statistics stabilize near ideal uniformity, and bell curve plots illustrate how initially skewed distributions flatten into near-perfect uniform distributions. Notably, this convergence is achieved even with modest configurations (\(N=4,5\)), underscoring both the practicality and scalability of the approach.

\medskip

Viewed from a physics perspective, QSQS serves as both a conceptual and operational framework: a system where true entropy arises from measuring two complementary observables under an uncertainty constraint. As a realization of QSQS, QPP-RNG effectively transforms unavoidable system-level fluctuations into cryptographically useful true randomness.

\medskip

In the quantum-safe era, this system-embedded TRNG helps close the entropy gap by providing true randomness directly inside cryptographic modules, thereby reducing dependence on external entropy sources and specialized hardware. The alignment between physical theory (QSQS and uncertainty) and engineering realization (QPP-RNG) points toward a unified, physics-informed foundation for future cryptographic systems.

\medskip

Future work will focus on formalizing the QSQS framework mathematically, exploring deeper connections to quantum information theory, and integrating QPP-RNG into advanced cryptographic protocols --- moving toward a fully self-contained, entropy-rich cryptographic architecture grounded in physical principles.

%\begin{acknowledgments}
%We wish to acknowledge the support of the author community in using
%REV\TeX{}, offering suggestions and encouragement, testing new versions,
%\dots.
%\end{acknowledgments}

\bibliography{my.bib}% Produces the bibliography via BibTeX.

\end{document}